\documentclass[%
reprint,
superscriptaddress,
amsmath,amssymb,
aps,floatfix, nofootinbib
]{revtex4-1}
\usepackage{graphicx}
\usepackage{bm}
\usepackage{braket}
\usepackage {colortbl,array,xcolor}
\usepackage{comment}
\usepackage{graphicx}
\usepackage{epsfig}
\usepackage{color}

\makeatletter

\@ifundefined{textcolor}{}
{%
 \definecolor{BLACK}{gray}{0}
 \definecolor{WHITE}{gray}{1}
 \definecolor{RED}{rgb}{1,0,0}
 \definecolor{GREEN}{rgb}{0,1,0}
 \definecolor{BLUE}{rgb}{0,0,1}
 \definecolor{CYAN}{cmyk}{1,0,0,0}
 \definecolor{MAGENTA}{cmyk}{0,1,0,0}
 \definecolor{YELLOW}{cmyk}{0,0,1,0}
}
\preprint{APS/123-QED}

\makeatother

\begin{document}
\title{
Quantum state transfer of angular momentum via single electron photo-excitation from a Zeeman-resolved light hole
 }
\author{K. Kuroyama,$^{1,\ \ast,\ \dagger}$  M. Larsson,$^{1,\ \ast}$ J. Muramoto,$^{1}$ K. Heya,$^{1}$ T. Fujita,$^{2}$ G. Allison,$^{3}$\\ S. R. Valentin,$^{4}$ A. Ludwig,$^{4}$ A. D. Wieck,$^{4}$ S. Matsuo,$^{1,\ \ddagger}$ A. Oiwa,$^{2}$ and S. Tarucha$^{1,3,\ \S}$}

\footnote[0]{\noindent $^*$ These two authors equally contributed to this work}\\
\footnote[0]{$^\dagger$ kuroyama@meso.t.u-tokyo.ac.jp}\\
\footnote[0]{$^\ddagger$ matsuo@ap.t.u-tokyo.ac.jp}\\
\footnote[0]{$^\S$ tarucha@ap.t.u-tokyo.ac.jp}\\

\footnote[0]{\noindent $^1$ Department of Applied Physics, The University of Tokyo, Bunkyo-ku, 7-3-1 Hongo, Tokyo, 113-8656, Japan\\
$^2$ The Institute of Scientific and Industrial Research, Osaka University, 8-1, Mihogaoka, Ibaraki-shi, Osaka 567-0047, Japan\\
$^3$ Center for Emergent Materials Science (CEMS), RIKEN, 2-1 Hirisawa, Wako-shi, Saitama, 351-0198, Japan\\
$^4$ Lehrstuhl f\"{u}r Angewandte Festk\"{o}rperphysik, Ruhr-Universit\"{a}t, Bochum, Universit\"{a}tsstr\"{a}e 150, D-44780 Bochum, Germany\\} 
 
\begin{abstract}
Electron spins in GaAs quantum dots have been used to make qubits with high-fidelity gating and long coherence time, necessary ingredients in solid-state quantum computing. The quantum dots can also host photon qubits with energy applicable for optical communication, and can show a promising photon-to-spin conversion. The coherent interface is established through photo-excitation of a single pair of an electron and a Zeeman-resolved light-hole, not heavy-hole. However, no experiments on the single photon to spin conversion have been performed yet. Here we report on single shot readout of a single electron spin generated in a GaAs quantum dot by spin-selective excitation with linearly polarized light. A photo-electron spin generated from a Zeeman-resolved light-hole exciton is detected using an optical spin blockade method in a single quantum dot and a Pauli spin blockade method in a double quantum dot. We found that the blockade probability strongly depends on the photon polarization and the hole state, heavy- or light-hole, indicating a transfer of the angular momentum from single photons to single electron spins. Our demonstration will open a pathway to further investigation on fundamental quantum physics such as quantum entanglement between a wide variety of quantum systems and applications of quantum networking technology. 
\end{abstract}

\maketitle
\onecolumngrid
\section*{}
\noindent{\bf Introduction.}
The concept of quantum state transfer (QST) between quantum particles of the same sort and different sorts arises from the universality of quantum mechanics, and has been demonstrated in media conversion from single photons to single atoms \cite{Kurz2014}, spins \cite{Yang2016}, for example. For applications in quantum information technology photons can be used as flying qubits as a way to network local quantum processors with solid-state qubits. GaAs semiconductor quantum dots (QDs) with g-factor engineering technique \cite{Allison2014} provide a promising solid-state platform hosting both single photons and single spins with an intrinsic interface that transferes the angular momentum \cite{Kosaka2003, Rao2005, Kuwahara2010, Pioda2011, Fujita2013}. Gate-defined GaAs QDs can suit the building blocks for quantum processing owing to fast control of electron spins \cite{Yoneda2014}, high-fidelity readout \cite{Nakajima2017}, possible scale-up of qubits \cite{ito2016} and a long memory time of milliseconds \cite{Malinowski2017}. 
\\

The photon to spin interface has been traditionally studied for semiconductor systems using optical techniques of e.g. circularly polarized light excitation of semiconductor systems based on selection rules \cite{Vrijen2001}. The coherence in this optical process has been demonstrated using a Kerr rotation technique that optically measures the spin orientation\cite{Kosaka2008}. The tomographic Kerr measurement was also performed for the Zeeman resolved light-hole (LH) excitons in GaAs quantum wells (QWs)\cite{Kosaka2009, Inagaki2014}. Experiments to directly detect spin information of the single photo-generated electrons in QDs are indeed a necessary ingredient for QST but a challenging task. 
\\

Previously, we have used optical selection rules in generating single photo-electrons and Pauli spin blockade (PSB) in detecting the spin orientation to verify angular momentum transfer from a single circularly polarized photon to a single electron spin in a laterally gate-defined GaAs QD \cite{Fujita2015}. A $\rm{}S_z=1/2\ (-1/2)$ conduction electron generated from a $\rm{}J_z=-3/2\ (3/2)$ valence electron by a $\rm{}\sigma^- (\sigma^+)$ photon remains in the dot, whereas the heavy-hole (HH) instantly escapes from the dot. Note that for quantum wells grown on a (001) substrate the HH state is not spin resolved by Zeeman energy, but the LH state is. Therefore, the superposition of of $\rm{}S_z=+1/2\ \mathrm{and}\ -1/2$ electron ($\ket{\rightarrow}\ \mathrm{and}\ \ket{\leftarrow}$) or spin qubit is only converted from the photon qubit in the photo-excitation from the spin-resolved LH state (see Eq. (1))
\\

Here we perform spin-selective photo-excitation of the single HH and LH excitons in a laterally gated GaAs QD, and use a charge sensing technique to measure the probability of finding single photo-electrons in the dot that are created in line with the optical selection rules which preserve the angular momentum. At first we use a single QD (SQD) having zero or one electron, $N_e\!=\!0$ or 1, and pump just one electron in the dot by linearly polarized light, vertical (V) or horizontal (H). We show that the photo-excitation of a pair of an electron and one of the spin-resolved LHs  is prohibited by Pauli exclusion in case the photo-generated spin is parallel to the residing electron spin in the $N_e\!=\!1$ SQD (optical spin blockade, OSB) \cite{Hogele2005}. Next we use a double QD (DQD) to detect the spin orientation, up or down, of the photo-generated electron in the spin-resolved LH excitation using PSB. In this experiment the DQD is formed by changing the gate voltage setting from that for the SQD, with having just one electron in one of the two dots, and one of the spin resolved-LH states is resonantly excited in the other dot. Finally we show that the obtained results are consistent between the two experiments. 
\\

\section*{}
\noindent{\bf Result}\\
\noindent{\bf Principles and expected results of photon-to-spin QST in LH state excitation.} In a Voigt geometry with an in-plane magnetic field, irradiation of linearly polarized photons selectively generates electrons whose spins are parallel ($\ket{\rightarrow}$) or anti-parallel ($\ket{\leftarrow}$) to the magnetic field. Therefore from a superposition state of two photon polarizations, $\ket{\psi}_{ph}=\alpha\ket{H}+\beta\ket{V}$ (H and V are a linear polarization which is perpendicular and parallel to the magnetic field, respectively) a simple product of a superposition state of $(\alpha\ket{\leftarrow}_e+\beta\ket{\rightarrow}_e)$ and a spin-resolved LH state, LH-: $\ket{\Leftarrow}_{lh}$ or LH+: $\ket{\Rightarrow}_{lh}$ (see Fig.\ref{fig1}(a) and (b)) is generated:
\begin{eqnarray}
\alpha\ket{H}+\beta\ket{V}\Longrightarrow\left(\alpha\ket{\leftarrow}_e+\beta\ket{\rightarrow}_e\right)\otimes\ket{\Leftarrow}_{lh} (\mathrm{or} \left(\alpha\ket{\leftarrow}_e-\beta\ket{\rightarrow}_e\right)\otimes\ket{\Rightarrow}_{lh} ).
\end{eqnarray}
We first study OSB in single photo-electron generation by selective photo-excitation of  electron and a HH or LH pairs in SQDs for $N_e\!=\!0$ or 1.  For the HH excitation the g-factor is 0 \cite{Jeune1999}, so both of the HH up-spin and down-spin states contribute to the optical transition. For the linear polarized photons, the spin configuration of the excited electron-HH pair is expressed as :
\begin{eqnarray}
\alpha\ket{H}+\beta\ket{V}&\Longrightarrow&\frac{1}{\sqrt{2}}\left(\alpha+\beta\right)\ket{\uparrow}_e\otimes\ket{\Uparrow}_{hh}+\frac{1}{\sqrt{2}}\left(\alpha-\beta\right)\ket{\downarrow}_e\otimes\ket{\Downarrow}_{hh}\nonumber\\
&=&\frac{1}{2}\left(\alpha+\beta\right)\left(\ket{\leftarrow}_e+\ket{\rightarrow}_e\right)\otimes\ket{\Uparrow}_{hh}+\frac{1}{2}\left(\alpha-\beta\right)\left(\ket{\leftarrow}_e-\ket{\rightarrow}_e\right)\otimes\ket{\Downarrow}_{hh}\nonumber\\
&=&\frac{1}{2}\ket{\leftarrow}_e\otimes\left[\left(\alpha+\beta\right)\ket{\Uparrow}_{hh}+\left(\alpha-\beta\right)\ket{\Downarrow}_{hh}\right]+\frac{1}{2}\ket{\rightarrow}_e\otimes\left[\left(\alpha+\beta\right)\ket{\Uparrow}_{hh}-\left(\alpha-\beta\right)\ket{\Downarrow}_{hh}\right]
\end{eqnarray}
For any values of $\alpha$ and $\beta$, the probability amplitude is 1/2 in all four terms of electron-hole pairs in Eq. (2). This holds for the HH excitation in the $N_e\!=\!0$ SQD. On the other hand for $N_e\!=\!1$ the electron ground state, up-spin state $\ket{\leftarrow}_e$ is already occupied, so that the optical transition expressed in the first square bracket of Eq. (2) does not occur. Therefore, the photo-electron trapping efficiency for $N_e\!=\!1$ is reduced to half the value for $N_e\!=\!0$.
\\

For the LH excitation in the $N_e\!=\!0$ SQD, which is represented by Eq. (1), the H (V)-polarization light generates an optical transition from LH- to $\ket{\leftarrow}_e$ ($\ket{\rightarrow}_e$) (see Fig. \ref{fig1}(a)). Therefore, the optical transition by the V-polarization light is forbidden for the $N_e\!=\!1$ SQD (see Fig.\ref{fig1}(b)). 
\\

Fig.\ref{fig1}(c), and (d) is a sketch of the HH and LH excitation spectrum by the V-, and H-polarization light, respectively.  The black, and red lines indicate the spectrum for the $N_e\!=\!0$, and 1 SQD, respectively.  The polarization dependence is opposite for the LH+ excitation, i.e., the transition by the H-polarization light is forbidden for $N_e\!=\!1$.
\\

Finally the expected spectrum of the optical transition probability subject to OSB is schematically shown in Fig.\ref{fig1}(c), and (d) for the V-polarization, and H-polarization light, respectively. Here the resonance peaks of the Zeeman split LH+, and LH- are indicated by the red, and blue line, respectively. Note that for $N_e\!=\!0$ the LH excitation probability is one third of the HH excitation probability reflecting the oscillator strengths.

\section*{}
\noindent{\bf Experimental techniques.} The QD device studied here is made in a two-dimensional electron gas (2DEG) accumulated in an AlGaAs/GaAs/AlGaAs QW. Two kinds of GaAs QW wafers are used to fabricate the QD devices (Supplementary note 1 and 2). The first wafer has a well width grading between 12 and 15 nm across a quarter of a 2 inch wafer and is used to fabricate the SQD for the OSB experiment.  Therefore the actual well width of the SQD depends on the wafer position used for the fabrication, and we roughly estimate the width of 13$\pm$0.5 nm. The DQD for the PSB experiment is fabricated using the second wafer. This wafer has a fixed well width of 15 nm. The QDs are defined by applying appropriate voltages to the surface gates and the quantum point contact (QPC) formed on the right of the dot by gate S was used as a charge sensor (see Fig.\ref{fig2}(a)). The QPC sensor is embedded in a radio-frequency (rf), impedance-matched circuit with resonance frequency of 214.5 MHz \cite{Barthel2010} which allows fast read-out of the photo-generated electrons. The tunnel coupling of the dot to the right lead was carefully adjusted with gates T, TR and R to be in the range of 0.2 to 20 kHz which is comparable to or lower than the charge sensor band width, while negligible to the left lead. A 100--to-200-nm-thick dielectric layer of calixarene is deposited on top of the central region of the device. A 300-nm-thick Ti/Au metal mask having a 500 nm diameter aperture is centered over the device. 
\\

Photon-trapping measurements were performed with the device placed in a dilution refrigerator at the base temperature of 25 mK with a window for optical access (See Fig.SI2). The single photo-electron trapping data was acquired in a single shot manner using the rf-QPC circuit. The photon trapping probability is determined from the number of photo-electron detection events as well as the incident photon flux that are calculated from the laser power and, aperture size. For photo-excitation we used pulsed photons generated from a wavelength tunable Ti:sapphire laser and a pulse picker combined with mechanical shutters for the OSB experiment with the SQD. The photons are irradiated onto the SQD through the aperture in the metal mask. For the PSB experiment with the DQD we used a weak continuous wave (CW) laser with an intensity of about 2.5 fW, equivalent to 10 kHz as photon rate (Supplementary note 3). 
\\

\noindent{\bf Single-shot read-out of a photo-generated electron.} We first tuned the SQD device to accumulate just a few electrons in the dot. Fig.\ref{fig2}(b) shows the charge state stability diagram measured with the charge sensor when sweeping the L and R gates. The diagonal lines indicate the charge state transitions. The charge number $N_e$ is fixed in each region of Coulomb blockade between the neighbouring lines. The charge transition line of $N_e\!=\!0$ to 1 appears jagged, because the dot-lead tunnel rate is lower than the gate voltage sweep rate. We set the gate bias point A (B) in the $N_e\!=\!0$ (1) state in Fig.\ref{fig2}(b) for the photon-trapping experiment.
\\

We performed the single shot read-out of the photo-generated electrons at point A in the  $N_e\!=\!0$ Coulomb blockade region by irradiating a photon pulse at t = 0 and simultaneously sampling the rf-QPC signal. The signal to noise ratio is unity for a 250 $\rm{}\mu s$ integration time on a sampling digitizer (AlazarTech ATS9440) with a microwave power of -75 dBm. The read-out speed is approximately 10 times higher than that in our previous measurement using a standard QPC charge sensing.
\\

Fig.\ref{fig2} (c) and, (d) show the typical photo-response, $\rm{}\Delta G_{sensor}$ of the charge sensor. In both figures we observe an abrupt change of $\rm{}\Delta G_{sensor}$ at t = 0, but the change is much larger and sharper in (d). The large photo-response in (d) is assigned to a single photo-electron trapping event on the dot, whereas the small photo-response in (c) is caused by an instant change of the charge configuration around the sensor, not by electron trapping on the dot.  In (d) the photo-generated electron finally tunnels out of the dot and $\rm{}\Delta G_{sensor}$ returns to the original value. The time resolution of $\rm{}\Delta G_{sensor}$ is 250 $\rm{}\mu sec$, shorter than the electron spin lifetime, and therefore we are able to detect the orientation of the photo-generated electron spin before the spin relaxes.
\\

\noindent{\bf Photo-luminescence spectrum of the QW wafer.} In order to evaluate the HH and LH excitation energy in the dot fabricated from the QW wafer we measured the photo-luminescence (PL) spectrum at 77 K in the absence of a magnetic field (see Fig.\ref{fig2}(e)). The temperature is relatively high so that both HH and LH excitons are populated. The PL spectrum is asymmetric due to the contribution from the LH excitons to the high energy side. The main peak is assigned to the HH exciton at 1.5276 eV, and the shoulder at higher energies is assigned to the LH exciton at 1.5341 eV. The exciton resonance energies are consistent with calculations for a one-dimensional finite well potential. The GaAs band gap increases with decreasing temperature and therefore, the HH and LH resonance peaks shift to higher energy by about 11 meV when the temperature is lowered to 0.1 K.
\\


\section*{}
\noindent{\bf Result of optical spin-blockade.}
 The first QW wafer used here is specially designed to study the OSB effect such that electron Zeeman energy is larger than the excitation light band width ($\rm\Delta\nu_{photon}$=0.6 meV) and the Zeeman splitting LH- or LH+ state is well resolved. The Zeeman energy estimated is 162 $\rm{}\mu$eV ($<\rm\Delta\nu_{photon}$) for electrons, assuming the g-factor of -0.4 \cite{LeJeune1997}, and 3 meV ($>\rm\Delta\nu_{photon}$) for LHs, assuming the g-factor of -3.5 \cite{Timofeev1996, Durnev2014} under a large in-plane magnetic field $\rm{}B_{//}$ = 7 T. This magnetic field is chosen to be large enough to polarize the electron spin but not so large as to reduce the spin lifetime below the readout time \cite{Fujita2015}. We finally find that the SQD device used here has large enough HH-LH separation to resolve the LH- state excitation for the photon-trapping experiment. 
\\

\begin{figure}[h]
\centerline{ \epsfig{file=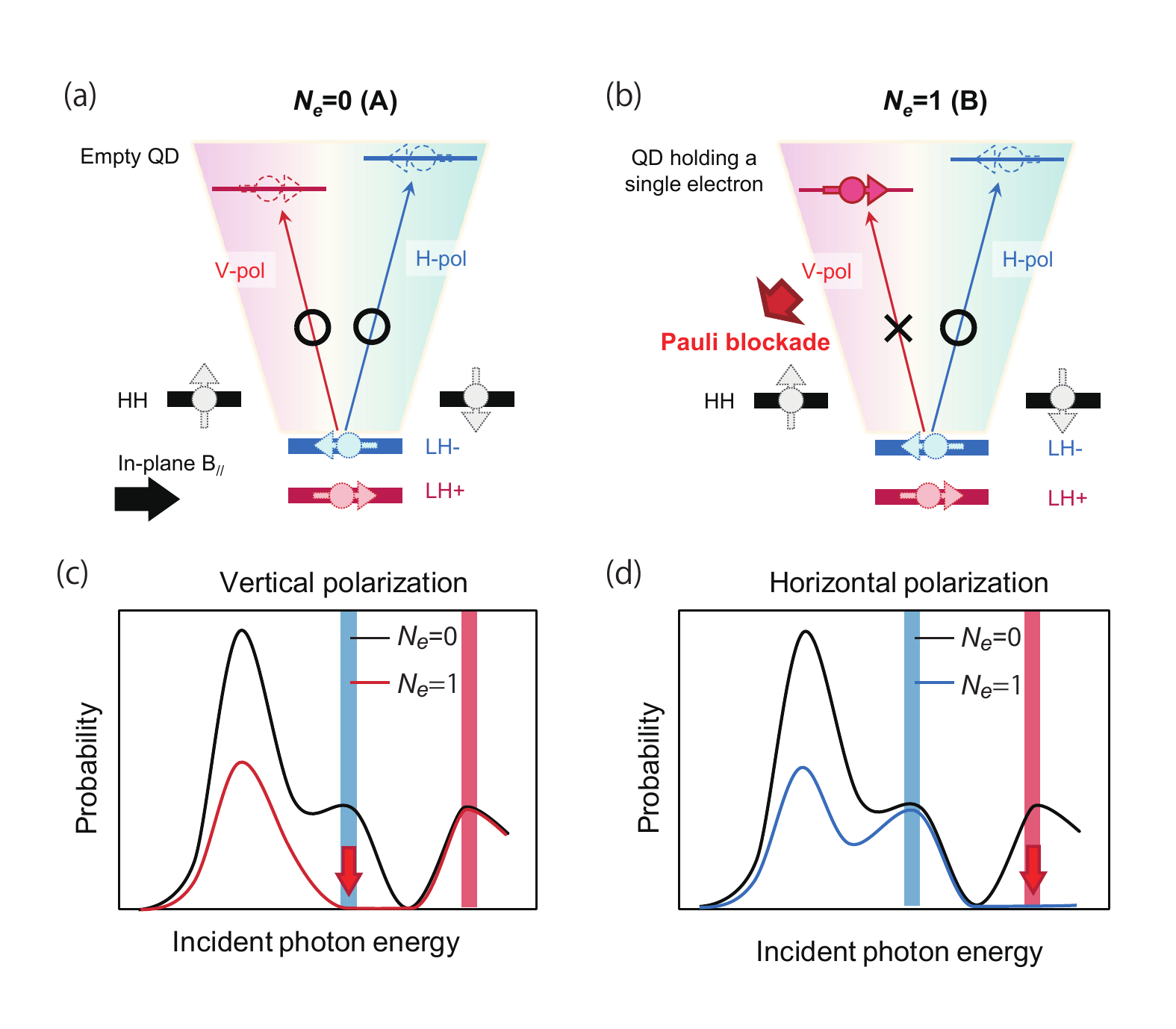,width=1\textwidth,clip}
} \caption{(Color online). (a), (b)Schematics of spin-selective optical transitions between the electron and hole states and the  optical spin blockade effect. For the $N_e\!=\!0$ dot, optical excitation of both an electron spin parallel and anti-parallel to an in-plane magnetic field is allowed. On the other hand, for the $N_e\!=\!1$ dot an electron spin anti-parallel to the magnetic field is initially trapped. In this configuration the optical transition with the V-polarized light is forbidden because of the Pauli exclusion rule in the dot. (c), (d) Energy spectrum of the photo-electron trapping probability in the dot expected from the principles of optical spin blockade. $N_e$ indicates the electron number initially trapped by the dot. The optical excitation with the V-polarized light from the lower Zeeman split LH state (highlighted in blue) is forbidden. On the other hand, for the upper Zeeman split LH state (highlighted in red) the excitation with the H-polarized light is forbidden.}
\label{fig1} 
\end{figure}

\begin{figure}[h]
\centering{ \epsfig{file=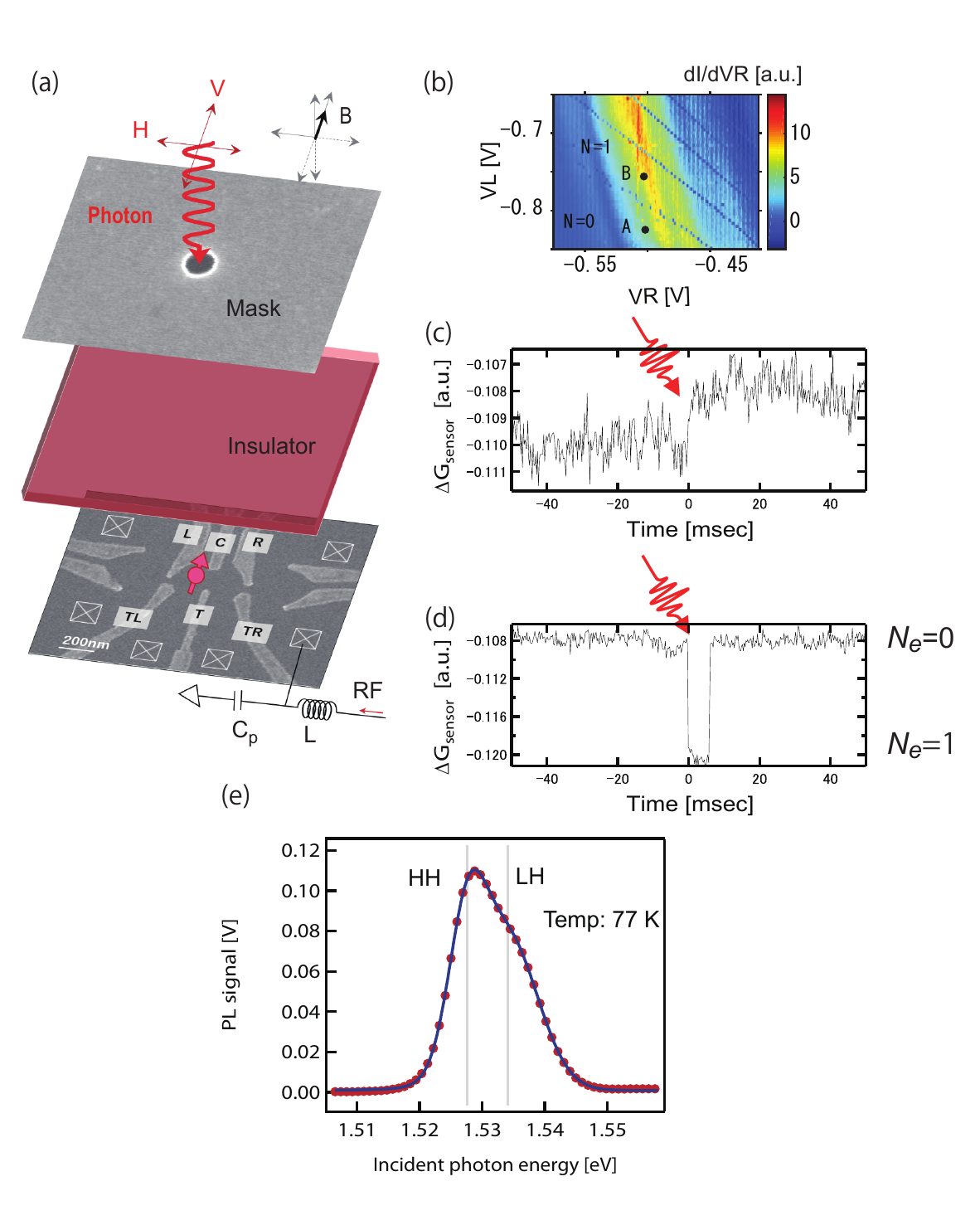,width=0.7\textwidth,clip}
} \caption{(Color online). (a) Schematic showing the device layout and directions of light polarization and external magnetic field. Photons are irradiated onto the dot passing through a 500 nm diameter aperture in a thick gold mask placed on top of the dot. (b) Charge stability diagram of the dot measured with the rf-QPC charge sensor. Point A, and B corresponds to the $N_e\!=\!0$ and $N_e\!=\!1$ charge state, respectively. The diagonal lines indicate the charge state transitions. (c), (d) Typical time traces of photo-response at the single QD. Both of the time traces are measured in the $N_e\!=\!0$ Coulomb blockade region and measured with the sampling rate of 250 $\rm{}\mu$sec. A pulsed photon is irradiated at t = 0 msec. (c) shows the conductance shift at the sensor QD solely due to the charge configuration change around it. On the other hand, in (d) $\rm{}\Delta G_{sensor}$ drops just after the photon irradiation and returns to the original level after some time, indicating a photo-electron is generated and trapped on the QD and then tunnels out. (e)Photo-luminescence spectrum of the wafer measured at 77 K without magnetic field. The left higher peak is assigned to HH excitation and the small shoulder located on the higher energy side of the HH peak is to LH excitation. }
\label{fig2} 
\end{figure}

\begin{figure}[h]
\centerline{ \epsfig{file=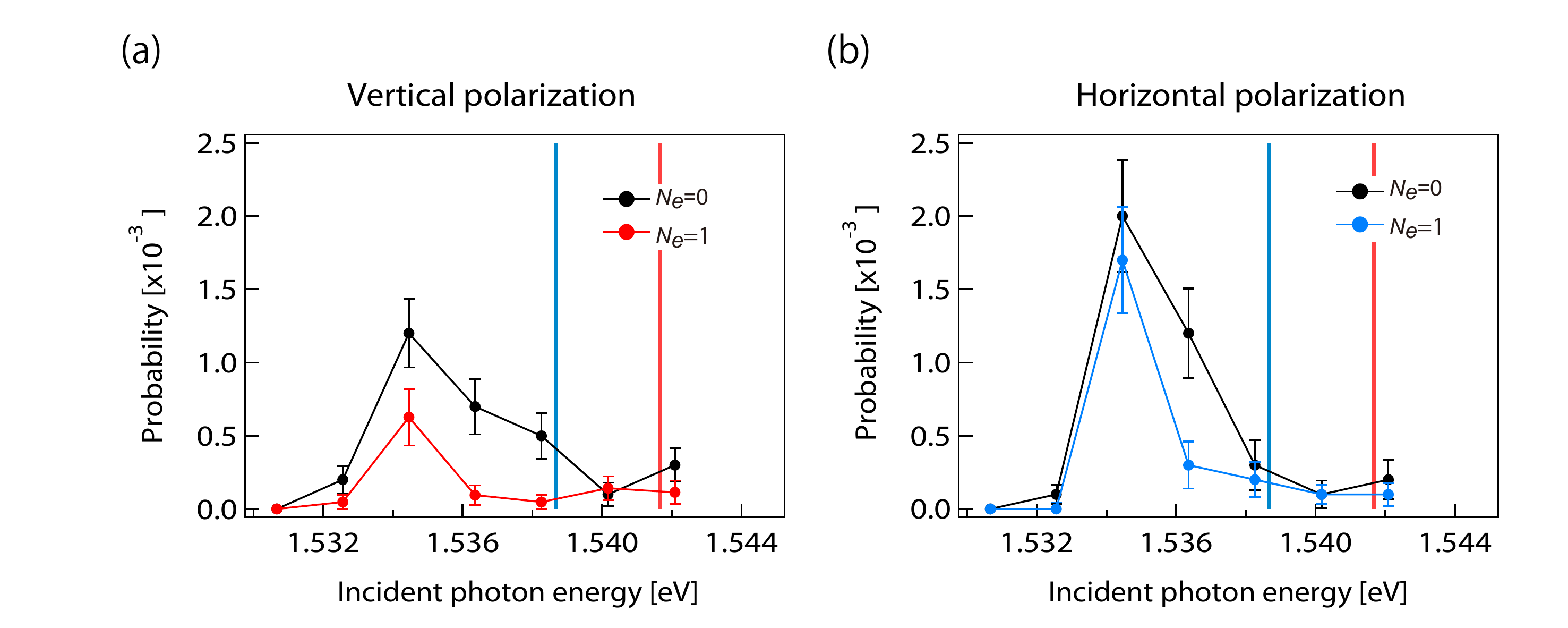,width=1\textwidth,clip}
} \caption{(Color online). Photo-electron trapping probability per photon passing through the aperture measured with the V-polarized light in (a) and H-polarized light in (b) as a function of excitation photon energy. Black plots indicate for $N_e\!=\!0$ and red (blue) plots for $N_e\!=\!1$ with the V(H)-polarized light. The magnetic field is 7 T. A peak at around 1.5345 eV indicates the HH state excitation. The vertical blue and red lines are LHs peak energies expected from previously reported values.}
\label{fig4} 
\end{figure}

The obtained photon-trapping probability spectra at $\rm{}B_{//}$ = 7 T, shown in Fig.\ref{fig4}(a) and Fig.\ref{fig4}(b), respectively, reproduce the OSB effect seen in Fig.\ref{fig1}(c), and (d), respectively. The estimate of electron temperature is less than 100 mK or 10 $\rm{}\mu$eV much smaller than the electron Zeeman energy, ensuring that $N_e\!=\!1$ QD is spin polarized up to 88 \%. Black closed circles, and colored closed circles in both figures represent the case for $N_e\!=\!0$, and 1, respectively. A peak at around 1.5345 eV observed for both $N_e\!=\!0$ and 1 and also for both light polarizations is assigned to the HH excitation. This peak energy is consistent with that predicted from the PL data in Fig. \ref{fig2}(d). No Zeeman splitting is observed for the HH exciton peak, while for $N_e\!=\!0$ in both figures, the photon trapping probability for the LH excitation shows a dip at around 1.540 eV due to the Zeeman splitting (Supplementary note 2). The blue, and red line indicates the calculated LH-, and LH+ exciton energy, respectively, using a value from literature of the LH exciton Zeeman energy \cite{Timofeev1996, Durnev2014}. We observe a feature, shoulder or kink, at the blue line, reflecting the LH- excitation and an increasing probability to the red line. Note the LH+ excitation may be significantly affected by the HH continuum excitation that is expected to be 7 to 8 meV above the HH exciton peak \cite{Bastard1982, Hrivnak1992}.  So here we focus on the OSB effect in the LH- excitation.   
\\

Now we compare the photon-trapping probability for $N_e\!=\!1$ to that of $N_e\!=\!0$ to reveal the OSB effect. For the V-polarization in Fig.\ref{fig4}(a) we observe a feature similar to Fig.\ref{fig1}(c), i.e., the HH peak is reduced by nearly half the LH- peak is strongly suppressed when $N_e$ is changed from 0 to 1. The LH- reduction is about 91 $\pm$ 9.8 \% probably limited by the electron spin polarization. On the other hand, for the LH+ excitation the reduction of $N_e\!=\!1$ is smaller, about 62 $\pm$ 30 \% inconsistent with that in Fig.\ref{fig1}(c), but probably due to the influence from the HH continuum excitation. 
\\

For the H-polarization in Fig.\ref{fig4}(b) the difference of the HH peak between $N_e\!=\!0$ and 1 is not so clear compared to that for the V-polarization, although qualitatively consistent with Fig.\ref{fig4}(a). The HH peak for $N_e\!=\!1$ is weakly reduced at the peak energy (by 85 $\pm$ 24 \%) but largely so on the high energy side (by more than 50 \%). The LH- peak is only slightly reduced, which is consistent with Fig.\ref{fig1}(d). The reduced probability for the LH+ excitation is inconsistent with Fig.\ref{fig1}(d) probably due to the overlap with the HH continuum as explained above.
\\

The $N_e\!=\!0$ photon-trapping probability spectrum, particularly for the HH excitation, is apparently different between the V- and H-polarization light excitation in Fig.\ref{fig4}(a) and (b) although they should be the same as illustrated in Fig.\ref{fig1}(c) and (d). The reason is not clear but it can be due to asymmetry in the optical coupling to the dot through the metal mask aperture between the V- and H-polarization light because the dot shape is elliptic. Nevertheless, we observe a significant influence of OSB on the photon trapping efficiency in the LH- excitation.
\\

\section*{}
\noindent{\bf Discussion}\\
\noindent{\bf Spin readout of a photo-electron using a double QD.}  In the preceding paragraphs we addressed the photon to spin conversion through the LH- excitation using the OSB effect in the SQD but not through the LH+ excitation because the LH+ peak is overlapped with the HH continuum excitation. We here used the second wafer having a slightly larger well width to fabricate the DQD, because the HH-LH separation is smaller and no large overlap of the LH+ and HH continuum excitation is assumed. Indeed we confirm that this assumption holds from measurements of the PL spectrum (see Fig.SI1(c)) and photon-trapping spectrum (See Fig.SI1(d) and (e)). We then apply a PSB spin readout technique for the LH+ excitation.
\\

The measurement method in detail is explained in \cite{Fujita2015}. Here we briefly summarize it. First, a DQD is prepared in the (0,1) state. When an electron with spin antiparallel to the magnetic field is generated in the left dot by the V-polarized photon, the photo-electron can tunnel to the right dot. Especially, when two singlet states of S(1,1) and S(0,2) are energetically aligned, the inter-dot electron tunneling of S(1,1)-S(0,2) transition repeatedly occurs. This is not the case for the H-polarized photon excitation, because a triplet state of T+(1,1) with two up spins is created and the inter-dot electron tunneling is blocked by Pauli exclusion. Therefore, the difference of the photo-generated spin configuration, parallel or anti-parallel, can be distinguished by a single-shot measurement of the charge change in the DQD (Supplementary note 4)\cite{Fujita2015}.Note that a photo-electron can be created in either dot of the DQD, and therefore we only post-selected the events of photo-electron trapping in the left dot to derive the probability of finding the parallel or antiparallel spin configuration.
\\

\begin{figure}[h]
\centerline{ \epsfig{file=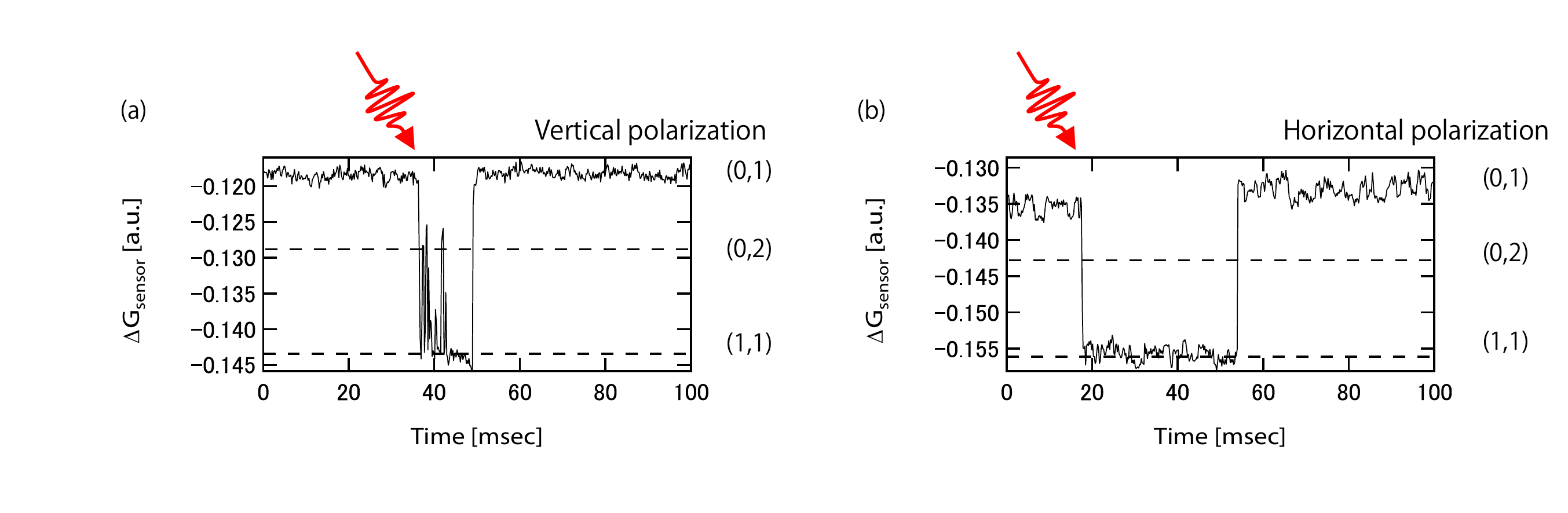,width=1\textwidth,clip}
} \caption{(Color online). Typical time traces of a photo-electron trapping on the DQD. (a) Photo-electron trapping signal with V-polarization photon excitation. The signal level of the charge sensor shows the oscillation between (0,2) and (1,1) charge states just after the photo-electron trapping, indicating that the photo-generated electron has a spin anti-parallel to the magnetic field. (b) For the case of the H-polarized photon excitation no oscillation is seen, because the electron spin is parallel to the prepared electron spin and blocked in the left dot by the Pauli exclusion rule.}
\label{fig5} 
\end{figure}

Fig.\ref{fig5} shows a typical charge sensor photo-response obtained for the V-, and H-polarized photon excitation in (a), and (b), respectively at in-plane magnetic field $\rm{}B_{//}$ = 7 T. We observe inter-dot oscillations between (1,1) and (0,2) upon the photo-electron trapping in (a) but not in (b). The probability of finding the inter-dot oscillation or antiparallel spin configuration is 53 \% for the V-photon excitation but 0 \% for the H-photon excitation. The probability for the H-excitation is as expected but that for the V-polarization is smaller than 100 \%. This is probably because of unintentional misalignment of the (1,1) and (0,2) states induced by the photon irradiation. Nevertheless, the obtained result is consistent with prediction about the coherent photon-to-spin conversion in the spin selective LH+ excitation.   
\\

\noindent{\bf Conclusion.} In conclusion single photon to single electron spin angular momentum transfer through spin-resolved light-hole state excitation was confirmed by single-shot spin measurement with OSB in a SQD and PSB in a DQD. We observed that the photo-electron trapping probability was strongly reduced for the V-polarized photon excitation of the $N_e\!=\!1$ SQD due to the OSB effect, and for the H-polarized photon excitation of the $N_e\!=\!1$ DQD due to the PSB effect. These results consistently show angular momentum preservation and are encouraging for further demonstration of single photon to spin quantum state transfer with QDs and of generation of quantum entanglement between different particles. This will in turn open a path to advanced quantum technology of quantum media conversion and quantum communication based on quantum teleportation.

\section*{}
\noindent{\bf Acknowledgement}
\\
This work was partially supported by Grant-in-Aid for Young Scientific Research (A) (No. JP15H05407), Grant-in-Aid for Scientific Research (A) (No. JP16H02204, No. JP25246005), Grant-in-Aid for Scientific Research (S) (No. JP26220710, No. JP17H06120), Grant-in-Aid for Scientific Research (B)  (No. JP18H01813) from Japan Society for the Promotion of Science, JSPS Research Fellowship for Young Scientists (No. JP16J03037), JSPS Program for Leading Graduate Schools (ALPS) from JSPS, Japan Society for the Promotion of Science (JSPS) Postdoctoral Fellowship for Research Abroad Grant-in-Aid for Scientific Research on Innovative Area, "Nano Spin Conversion Science" (No.26103004), Grant-in-Aid for Scientific Research on Innovative Area, CREST (No. JPMJCR15N2). A.D.W., A.L. and S.R.V acknowledge gratefully support of DFG-TRR160, BMBF-Q.com-H 16KIS0109, and the DHF/UFA CDFA-05-06.
\\

\section*{}
\noindent{\bf Method}
\\
\noindent {\bf General Properties of our QD sample.} Our measured sample is fabricated from a AlGaAs/GaAs/AlGaAs double heterostructure wafer. Ti/Au Schottky gates which define QDs and the nearby charge sensor are fabricated by using electron beam lithography. The fine gates are covered with a 300 nm thick calixarene insulator, and the optical mask which has a 500 nm diameter aperture is placed on top of the insulator. The optical setup for photon irradiation is explained in SI. 
\\

\normalsize
\def\bibsection{}
\noindent{\bf References}
%


\newpage

\newpage 

\newpage

\newpage
\newpage

\title{Supplementary Information}
\maketitle
\section*{}
\clearpage

\section*{Supplementary Note 1: properties of QW wafers}
\noindent 
Fig.SI\ref{fig:figS1}(a) and (b) show the layer structures of two different QW wafers used to fabricate the SQD for the OSB experiment in (a) and the DQD for the PSB experiment in (b). The difference in the QW between the two wafers are explained in the main text. A distributed Bragg reflector is embedded below the QW in order to enhance photon absorption in the QW in both wafers. A two-dimensional electron gas (2DEG) is accumulated in the GaAs QW. The electron density is $-1.91\times10^{11} \mathrm{cm^{-2}}$ and the mobility is $1.24\times10^{6} \mathrm{cm^2/Vs}$ for the first wafer. They are $-1.91\times10^{11} \mathrm{cm^{-2}}$ and $2.19\times10^{5} \mathrm{cm^2/Vs}$ for the second.
\\

\noindent 
Fig. SI1(c) shows the photo-luminescence (PL) spectrum of the second wafer measured at 77 K in the absence of a magnetic field. The temperature is relatively high so that the HH and LH excitons are both populated. The PL spectrum is asymmetric due to the contribution from the LH excitons to the high energy side. The main peak is assigned to the HH exciton at 1.5192 eV, and the shoulder on the high energy is also assigned to the LH exciton at 1.5236 eV. The HH and LH exciton energy are lower than those of the first wafer (see Fig. 2(e)) because the well is slightly wider in the the second wafer.
\\

\section*{Supplementary Note 2: Selection of the second QW wafer for the LH+ excitation}
\begin{figure}[h]
\begin{center}
\includegraphics[width=15cm]{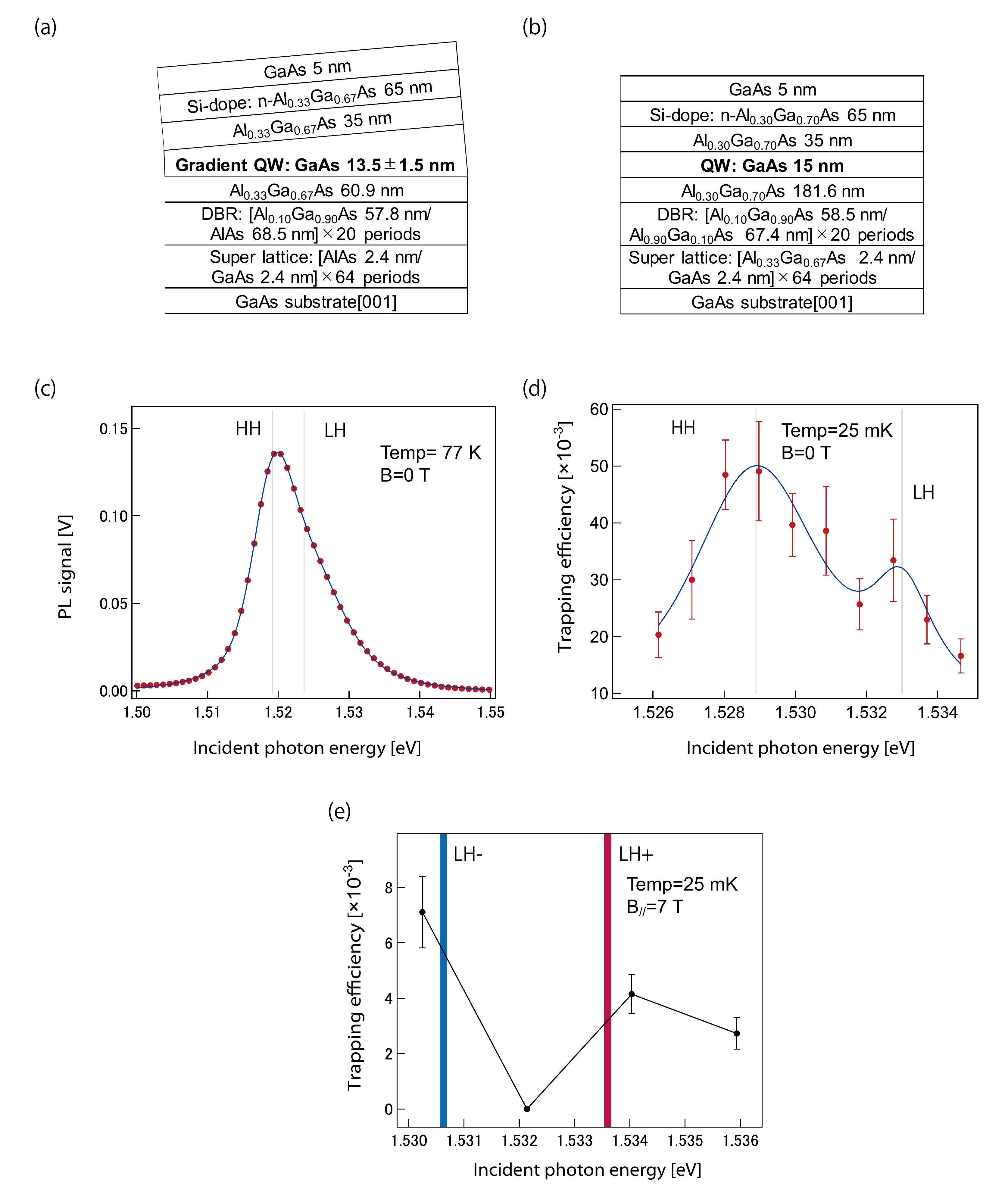}
\caption*{
\begin{flushleft}
\textbf{Figure SI1} (a)(b) Layer structure of the QW wafers. The GaAs QW is defined by a GaAs/AlGaAs heterostructure on top and bottom of the GaAs layer. The first wafer has a gradient QW layer, our sample is fabricated from the thinner region, so that the electron and a hole confinement of the first wafer is stronger than that of the second wafer. To enhance photon absorption at the QW, a distributed Bragg reflector (DBR) is embedded bellow the QW. (c) Photo-luminescence spectrum of the second wafer (Fig.(b)) measured at 77 K without magnetic field. The left higher peak is assigned to HH excitation and the small peak located on the higher energy side of the HH peak is to LH excitation. (d)Energy spectrum of photon-trapping efficiency measured with un-polarized light at $\rm{}B_{//}=0 T$ The HH and LH peaks are well distinguished. The blue curve is a fitting function constructed by two Voigt functions. (e)Photon-trapping efficiency around the LH peak obtained in (d). The LH peak split into two separated by Zeeman energy of approximately 3 meV. The blue and red bars are the locations of the LH Zeeman levels, as expected by previous results \citep{Timofeev1996, Durnev2014}.
\end{flushleft}
}
\label{fig:figS1}
\end{center}
\end{figure}

\noindent We used the SQD made from the first wafer for the OSB spin readout experiment (see main text). On the other hand we used the DQD made from the second wafer for the PSB spin readout experiment with the LH+ excitation, because we found that the LH+ excitation is not influenced from the HH continuum excitation in the DQD device whereas it is in the SQD device. During this selection we used non-polarized light to measure the photon trapping probability on a SQD device made from the second wafer to characterize the HH and LH state excitation and Zeeman splitting of the LH state excitation (Fig.SI\ref{fig:figS1}(d) and (e)). The measurement error is derived from the standard deviation of a binomial distribution of the measurement outcome.  In (d) we observe two peaks due to the HH and LH excitons with energy separation of 4.1 meV measured in the absence of a magnetic field. The photon trapping probability is approximately three times larger for the HH exciton than for the LH exciton, reflecting the difference of the HH and LH oscillator strengths \citep{Morimoto2014}. In (e) the LH state is separated into two, LH-, and LH+ state in the low, and high energy side, respectively by a Zeeman energy of about 3 meV at $\rm{}B_{//} =$ 7 T. It is clear that the photon trapping probability is strongly suppressed at the LH peak position observed at $\rm{}B_{//}$ = 0 T and becomes large at the LH-, and LH+ exciton positions (indicated by blue, and red line, respectively). We note that the LH+ excitation 5.6 meV above the HH peak is certainly not influenced from the HH continuum excitation, because we observe no photo-excitation at least 5.6 meV above the HH excitation in the SQD device made using the second wafer (see Fig. 3). Note that the HH-LH separation depends significantly on the well width but the Zeeman energy does not.  The calculated HH-LH separation is smaller by 1.9 meV for the 15 nm well than that of the 13 nm well.
\\

\section*{Supplementary Note 3: Optical setup for photon irradiation}
\begin{figure}[h]
\begin{center}
\includegraphics[width=17cm]{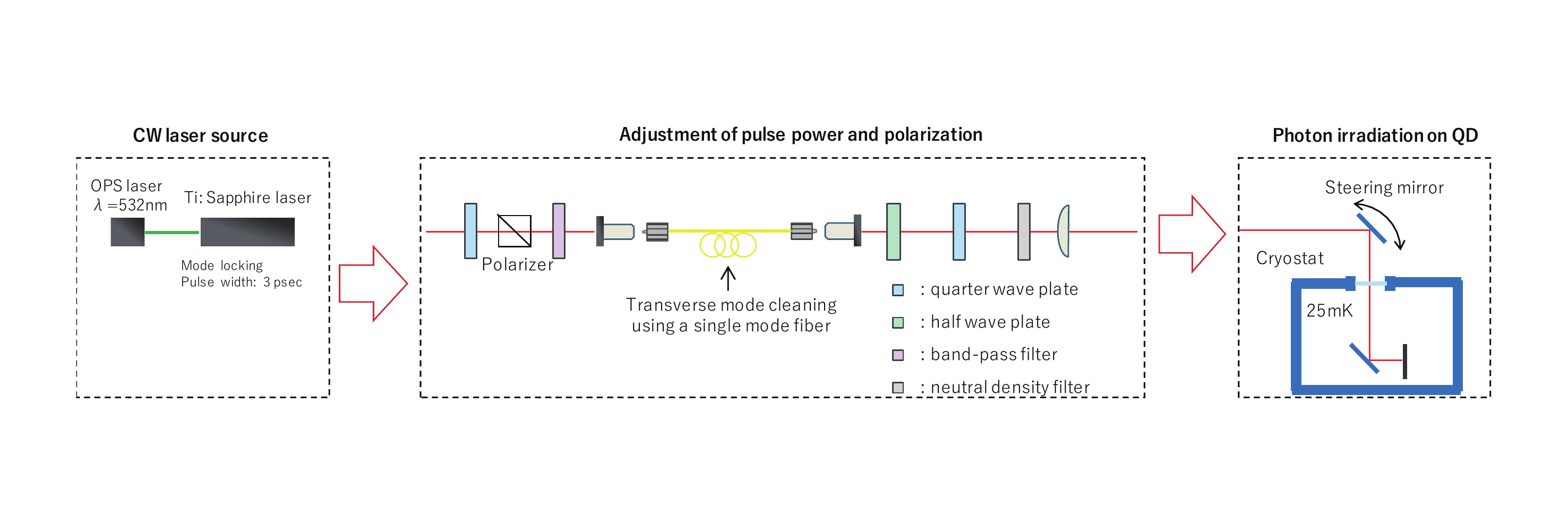}
\caption*{
\begin{flushleft}
\textbf{Figure SI2} Optical setup for the pulsed photon irradiation of the QD sample placed in the cryostat. A single pulse which is synchronized with an external trigger is extracted using the pulse picker and mechanical shutters. By transmitting the pulse into a single mode optical fiber, the TEM00 component of the pulsed laser remains. The polarization is adjusted to H or V using a polarizer and half wave plates, and the laser power is modulated by a neutral density filter. The laser spot is carefully aligned onto the QD sample by tilting the steering mirror.  
\end{flushleft}
}
\label{fig:figS2}
\end{center}
\end{figure}

\noindent For the OSB experiment, we used a pulsed laser source with an AOM pulse picker, and an overview of our setup is shown in Fig.SI\ref{fig:figS2}.  The pulse picker and mechanical shutters are synchronized with an external trigger, and once they are triggered, pick up a single pulse from a pulse train which is emitted from Ti; Sapphire laser with repetition rate of 76 MHz. In the second part of Fig.SI\ref{fig:figS2}, first, TEM00 component of a single pulse is extracted by using a single mode optical fiber. Afterwards photon polarization is prepared into a certain orientation using a polarizer, and the transmission at the polarizer is maximized by rotating a half wave plate (HWP). we prepared photon polarization of H or V using the second HWP. The pulse power is reduced with a neutral density filter such that only a few tens of photons are irradiated on the QD during a single pulse irradiation. In the final part, the laser spot position is precisely adjusted on the QD placed in the cryostat by titling the steering mirror.
\\

\section*{Supplementary Note 4: PSB on DQD for photo-electron spin readout}
\begin{figure}[h]
\begin{center}
\includegraphics[width=17cm]{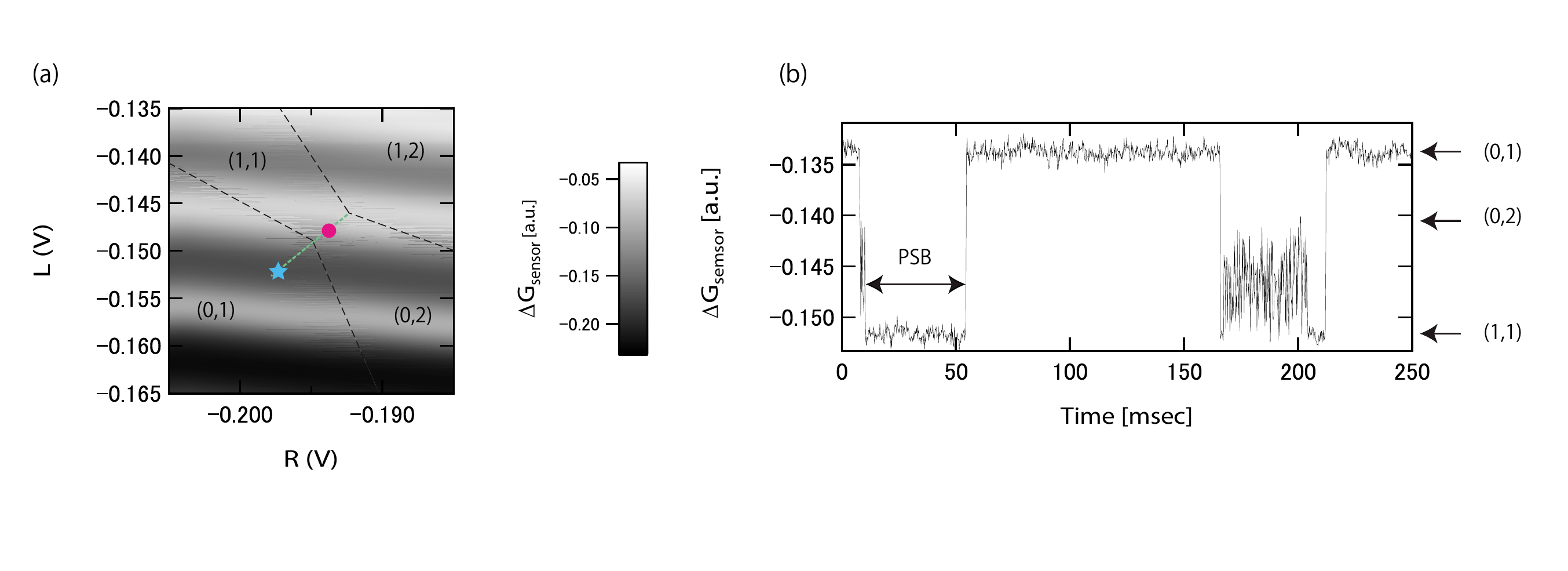}
\caption*{
\begin{flushleft}
\textbf{Figure SI3} (a) A stability diagram of the DQD sample at an in-plane magnetic field of 7 T. The shown region is around the resonance between the (1,1) and (0,2) charge states. The white and black oscillations in the background are Coulomb peaks in the QD charge sensor which is formed on the left side of the DQD. Black broken lines indicates a charge state transition, and the green is the resonance line of the (1,1) and the (0,2) states. We checked the inter-dot tunneling rate at the red circle point, and photon irradiation is done at the condition indicated by the blue star.(b) A typical time trace of inter-dot oscillation measured at the red point indicated in the Fig.(a). An electron is injected from the outer lead and shows inter-dot tunneling while two electron spins are anti-parallel to each other. After some time the oscillation is stopped at (1,1) charge state, indicating that the electron spins decay from S to $\rm{}T_+$.
\end{flushleft}
}
\label{fig:figS3}
\end{center}
\end{figure}

\noindent For spin readout of a photo-electron spin using Pauli spin blockade (PSB), we tuned the plunger gates called L and R (see Fig. 2(a)) such that (1,1) and (0,2) charge states are energetically aligned. The DQD charge stability diagram around the resonance of the (1,1) and (0,2) states is shown in Fig.SI\ref{fig:figS3}(a). The charge sensor QD is located on the left side of the dots, and the oscillations in the background of the diagram are Coulomb oscillations in the QD sensor. There are charge state transition lines between the neighbouring charge states, and the boundary of the (1,1) and (0,2) charge states indicates singlet resonance of S(1,1) and S(0,2) state. Therefore, when the gate voltages are tuned onto the red point, resonant transition of S(1,1) and S(0,2) is observed , and its real time trace is shown in Fig.SI\ref{fig:figS3}(b). The transition repeats until the spin state is relaxed to a ground state of triplet $\rm{}T_+(1,1)$ because of spin-orbit interaction \cite{Maisi2016, Fujita2016}.     
\\

Finally, for photo-electron spin readout we prepared the DQD in the (0,1) charge state region, but maintain the resonance of S(1,1)-S(0,2) above Fermi energy of source-drain leads. Note that the prepared electron spin is polarized to the magnetic field orientation, and therefore we can judge orientation of a photo-generated electron spin along the field by whether the resonant transition is observed between S(1,1)-S(0,2) just after a photon-trapping. 
\newline

\normalsize
\def\bibsection{}
\bibliographystyle{apsrev4-1}
\noindent{\bf References}
%


\newpage

\end{document}